\documentclass[
    ,final            
  ]
  {aipproc}

\layoutstyle{8x11double}

\begin{document}

\title[Zeeman]{Anisotropic Zeeman Splitting In Ballistic One-Dimensional Hole Systems}


\author{R. Danneau}{address={School of Physics, University of
New South Wales, Sydney, New South Wales, 2052, Australia.}}
\author{O. Klochan}{
address={School of Physics, University of New South Wales, Sydney,
New South Wales, 2052, Australia.}}
\author{W. R. Clarke}{
address={School of Physics, University of New South Wales, Sydney,
New South Wales, 2052, Australia.}}
\author{L. H. Ho}{
address={School of Physics, University of New South Wales, Sydney,
New South Wales, 2052, Australia.}}
\author{A. P. Micolich}{
address={School of Physics, University of New South Wales, Sydney,
New South Wales, 2052, Australia.}}
\author{M. Y. Simmons}{
address={School of Physics, University of New South Wales, Sydney,
New South Wales, 2052, Australia.}}
\author{A. R. Hamilton}{
address={School of Physics, University of New South Wales, Sydney,
New South Wales, 2052, Australia.}}
\author{M. Pepper}{
address={Cavendish Laboratory, Madingley Road, Cambridge, CB3 OHE,
United Kingdom.}}
\author{D. A. Ritchie}{
address={Cavendish Laboratory, Madingley Road, Cambridge, CB3 OHE,
United Kingdom.}}
\author{U. Z\"{u}licke}{address={Institute of Fundamental Sciences and MacDiarmid
Institute for Advanced Materials and Nanotechnology, Massey
University, Palmerston North, New Zealand.}}

\begin{abstract}

We have studied the effect of an in-plane magnetic field $B$ on a
one-dimensional hole system in the ballistic regime created by
surface gate confinement. We observed clearly the lifting of the
spin degeneracy due to the Zeeman effect on the one dimensional
subbands for $B$ applied parallel to the channel. In contrast, no
Zeeman splitting is detected for $B$ applied perpendicular to the
channel, revealing an extreme anisotropy of the effective
Land\'{e} \emph{g}-factor $g^{*}$. We demonstrate that this
anisotropy is a direct consequence of the one-dimensional
confinement on a system with strong spin-orbit coupling.

\end{abstract}

\maketitle

It has been proposed to exploit the intrinsic coupling between the
spin and orbital motion of quantum particles to control
spin-splitting with an electric field \cite{byshkov1984}, offering
new opportunities to implement a spintronic paradigm
\cite{zutic2004}. The ability to control the spin-splitting with
an external electric field has led to proposals for a
spin-field-effect transistors \cite{datta1990} which could be used
to perform new kinds of logic operations.

\begin{figure}[htbp]
\includegraphics[height=0.425\textheight]{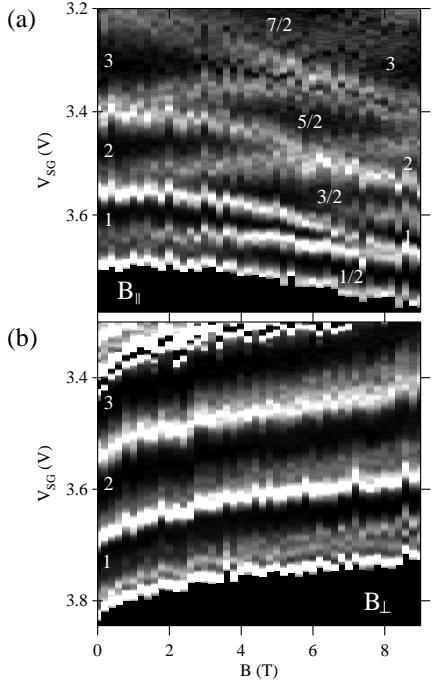}
\caption{Zeeman effect in a 1D hole system: (a) Transconductance
grayscale as a function of $V_{\mathrm{SG}}$ and $B_{\parallel}$;
black regions correspond to low transconductance (\emph{i.e.}
conductance plateaus, labelled in units of $2e^{2}/h$), white
regions correspond to high transconductance (subband edges):
splitting of the 1D subbands is clearly observed. (b)
Transconductance grayscale as a function of $V_{\mathrm{SG}}$ and
$B_{\perp}$: no Zeeman splitting is seen. Data are taken with a
back gate and mid-line gate voltage of 2.5 and -0.225 V
respectively (see \cite{danneau2006apl} for description of the
device).}
\end{figure}

Electric-field-tuneable spin-orbit (SO) interactions for electrons
arise through the coupling between conduction and valence band
that is induced by structural inversion asymmetry. Consequently,
such SO effects are larger in narrow-gap materials such as InGaAs
and InAs. However, the small band gap  makes it difficult to use
conventional metal surface gate technique, due to the absence of a
Schottky barrier, to create nanostructures for electrical field
control of spin studies. An alternative is to use holes rather
than electrons in a wider gap material. In a tight binding view,
as valence-band states are predominantly p-like (unlike
conduction-band states which are s-like), SO effects are
particularly important in low-dimensional hole systems such as
\emph{p}-type GaAs, making this material system interesting for
spin-controlled devices.

In bulk, zinc-blende compounds such as GaAs, SO coupling causes
splitting of the top-most valence band in four fold degenerate
heavy and light hole state (with total angular momentum $J = 3/2$)
and the two fold degenerate split off state (with total angular
momentum $J = 1/2$). However, two dimensional (2D) confinement
lifts the heavy hole (HH)-light hole (LH) degeneracy at $k=0$. As
a result the carrier transport is predominantly through the HH
subband. In addition, the confinement gives rise to mixing
($k\neq0$) and non-parabolicity of the HH and LH subbands (for a
complete review see \cite{winkler2003}).

We have studied how confining holes in a 1D channel affects their
spin properties. We used the intrinsic conductance quantization
properties of ballistic one-dimensional (1D) systems to probe the
1D subband edges and studied the effect of an in-plane magnetic
field $B$. We have performed Zeeman splitting measurements in a
ballistic 1D hole system aligned along the $[\overline{2}33]$
direction and formed in a GaAs (311)A quantum well by surface gate
confinement (see \cite{danneau2006apl} for description of the
device). Application of an in-plane magnetic field parallel
($B_{\parallel}$) to the wire lifts the spin degeneracy and
eventually causes the 1D subbands edges to cross. Figure 1 clearly
shows the splitting of the subband edges (white regions) in the
transconductance grayscale plot
($\mathrm{d}G/\mathrm{d}V_{\mathrm{SG}}$ as a function of side
gate voltage $V_{\mathrm{SG}}$ and $B$; the derivative has been
numerically calculated from the differential conductance $G$
corrected for a series resistance). After thermal cycling and
sample re-orientation, we have studied the effect of an in-plane
magnetic field perpendicular to the wire $B_{\perp}$. In contrast,
the transconductance grayscale (Fig.\ 1(b)) shows that the
degenerate 1D subbands are not affected by $B_{\perp}$ up to 8.8
T, \emph{i.e.} no Zeeman splitting is seen when the magnetic field
is aligned perpendicular to the channel.

Combining Zeeman effect measurements and source-drain bias
spectroscopy, and knowing that the spin-splitting is linear in $B$
(see \cite{danneau2006prl}) , one can extract the effective
Land\'{e} \emph{g}-factor $g^{*}$ using the basic relation $\Delta
E_{\mathrm{N}}=g^{*}_{\mathrm{N}}\mu_{\mathrm{B}}B$
\cite{fang1968}. The $g^{*}$ ratio
$g^{*}_{\parallel}$/$g^{*}_{\perp}$ (\emph{i.e.} for
$B_{\parallel}$ and $B_{\perp}$) can be estimated \emph{at least}
to be 4.5 \cite{danneau2006prl}, significantly larger than the
anisotropy calculated \cite{winkler2000} and measured
\cite{papadakis2000} in 2D hole systems. One can explain this
strong anisotropy by the following arguments. As mentioned above,
only HH participate in the carrier transport in a 2D hole system.
It has been demonstrated that Zeeman splitting is suppressed for a
magnetic field applied parallel to a 2D hole quantum well, in
which the total angular momentum axis \emph{{J}} is aligned along
the growth axis \cite{vankeresten1990,winkler2003}. Only a
magnetic field applied along \emph{{J}} generates the energy
splitting due to the Zeeman effect for both HH and LH bands.
\emph{This is a result of strong SO coupling in the valence band}.
In our system, the 1D confinement forces \emph{{J}} to lie along
the 1D constriction: this explains why no Zeeman splitting is
measured for $B$ applied perpendicular to the channel (see Fig.
2), though $B$ applied parallel to the channel lifts the spin
degeneracy of the 1D subbands.

\begin{figure}
\includegraphics[height=0.25\textheight]{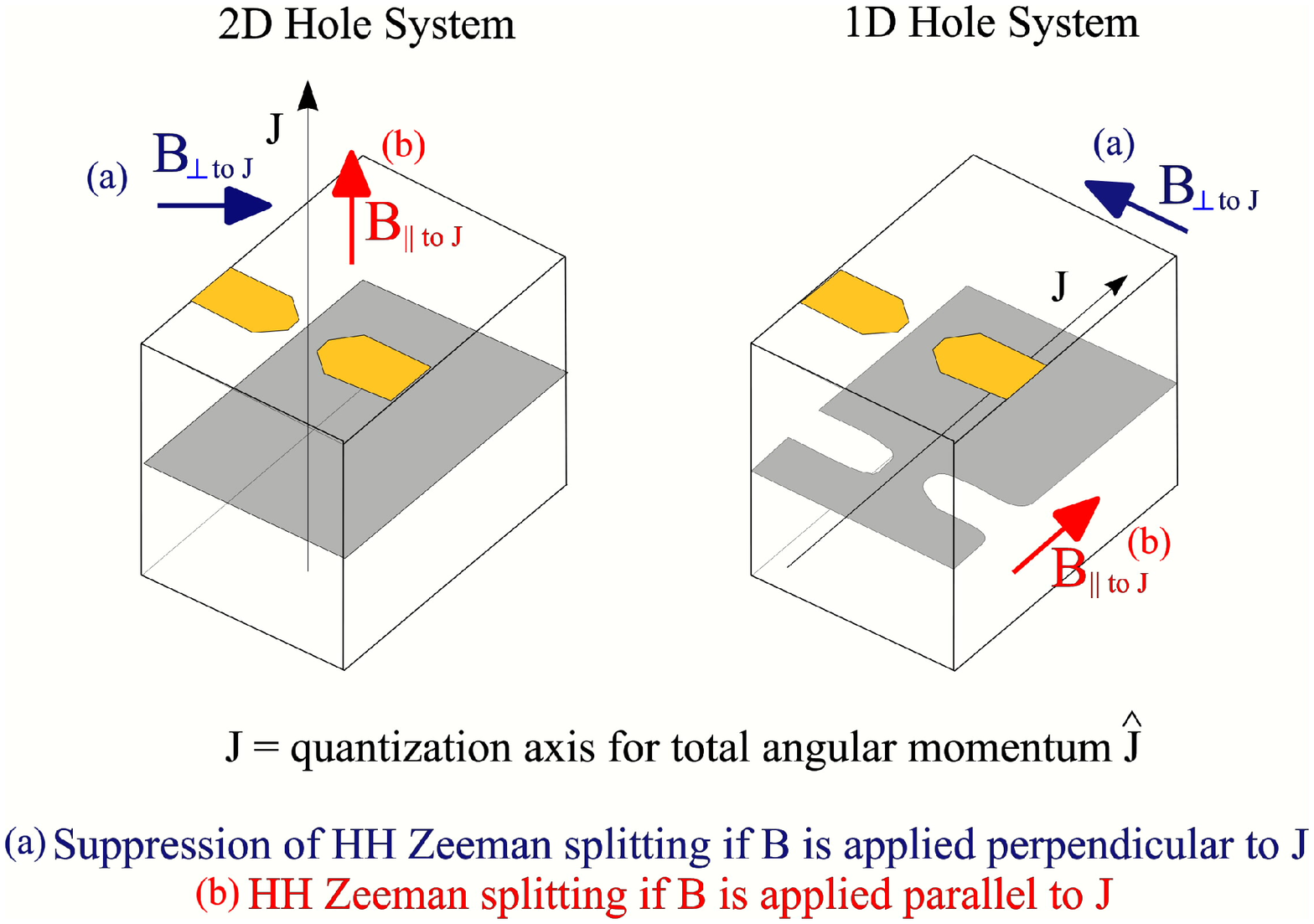}
\caption{Sketch of the effect of a magnetic field perpendicular
(a) and parallel (b) to the quantization axis for total angular
momentum \^{J}, on a 2D hole (as in \cite{vankeresten1990}) and 1D
hole system created by two side gates \cite{danneau2006prl}.}
\end{figure}

In conclusion, we have measured a strong anisotropy of the Zeeman
splitting in a 1D hole system with respect to an in-plane magnetic
field $B$ oriented along or parallel to the channel. Our results
show that confining holes to a 1D system fundamentally alters
their spin properties, and that it is possible to tune the $g^{*}$
anisotropy (as well as the absolute value of the effective
Land\'{e} $g$-factor \cite{danneau2006prl}), by electrostatically
changing the width of the 1D system.

We acknowledge support from the Australian Research Council, the
EPSRC and the Marsden Fund of the Royal Society of New Zealand. R.
D. acknowledges an Australian Research Council Postdoctoral
Fellowship and M. Y. S. acknowledges an Australian Research
Council Fedaration Fellowship.

\end{document}